\documentstyle[prd,aps]{revtex}

\newcommand{\bea}{\begin{eqnarray}}
\newcommand{\eea}{\end{eqnarray}}

\begin{document}

\draft
\twocolumn[\hsize\textwidth\columnwidth\hsize\csname
@twocolumnfalse\endcsname

\title{Third-order perturbations of a zero-pressure cosmological medium: \\
       Pure general relativistic nonlinear effects}

\author{Jai-chan Hwang${}^{(a)}$ and Hyerim Noh${}^{(b)}$}
\address{${}^{(a)}$ Department of Astronomy and Atmospheric Sciences,
                    Kyungpook National University, Taegu, Korea \\
         ${}^{(b)}$ Korea Astronomy and Space Science Institute,
                    Daejon, Korea
         }
\maketitle

\begin{abstract}

We consider a general relativistic zero-pressure irrotational
cosmological medium perturbed to the third order. We assume a flat
Friedmann background but include the cosmological constant. We
ignore the rotational perturbation which decays in expanding phase.
In our previous studies we discovered that, to the second-order
perturbation, except for the gravitational wave contributions, the
relativistic equations {\it coincide exactly} with the previously
known Newtonian ones. Since the Newtonian second-order equations are
fully nonlinear, any nonvanishing third and higher order terms in
the relativistic analyses are supposed to be pure relativistic
corrections. In this work we derive such correction terms appearing
in the third order. Continuing our success in the second-order
perturbations we take the comoving gauge. We discover that the
third-order correction terms are of $\varphi_v$-order higher than
the second-order terms where $\varphi_v$ is a gauge-invariant
combination related to the three-space curvature perturbation in the
comoving gauge; compared with the Newtonian potential we have
$\delta \Phi \sim {3 \over 5} \varphi_v$ to the linear order.
Therefore, the pure general relativistic effects are of {\it
$\varphi_v$-order higher} than the Newtonian ones. The corrections
terms are {\it independent} of the horizon scale and depend only on
the linear order gravitational potential (curvature) perturbation
strength. {}From the temperature anisotropy of cosmic microwave
background we have ${\delta T \over T} \sim {1 \over 3} \delta \Phi
\sim {1 \over 5} \varphi_v
 \sim 10^{-5}$.
Therefore, our present result reinforces our previous important
practical implication that near current era one can use the
large-scale Newtonian numerical simulation more reliably even as the
simulation scale approaches near (and goes beyond) the horizon.

\end{abstract}

\vskip2pc]

%
%
\section{Introduction}

In our previous works \cite{NL,second-order} we have proved that in
the zero-pressure irrotational cosmological medium the relativistic
second-order scalar-type perturbation equations {\it coincide
exactly} with the known ones in Newtonian theory. This result shows
a continuation of the relativistic-Newtonian correspondences of the
cosmological zero-pressure medium previously shown in the background
world model by Friedmann in 1922 \cite{Friedmann-1922} and by Milne
in 1934 \cite{Milne-1934}, and in the linear perturbation by
Lifshitz in 1946 \cite{Lifshitz-1946} and by Bonnor in 1957
\cite{Bonnor-1957}. History shows that both for the background and
for the linear perturbation the equations were first derived in
Einstein's gravity \cite{Friedmann-1922,Lifshitz-1946} and later
followed by studies in Newton's gravity
\cite{Milne-1934,Bonnor-1957}. In the case of second-order
perturbations the Newtonian result was known first
\cite{Peebles-1980}. Since the Newtonian results are not supposed to
be reliable as the scale approaches the horizon, our result has a
practical importance by showing that even to the second-order
perturbations we can use Newtonian equations in {\it all} scales. In
this work we will extend the situation to the next order in
relativistic perturbation which, as we will explain shortly, can be
regarded as pure relativistic corrections. Even in the linear
perturbations the presence of the tensor-type perturbation
(gravitational waves) can be regarded as a pure relativistic effect.
To the second order, additionally, the gravitational waves and the
scalar-type perturbation are coupled, thus one can source the other.
This can be regarded as another pure relativistic correction.

In this work we will derive pure relativistic corrections of the
scalar-type perturbation which appear in the third-order
perturbations. The situation we have is that to the second order
Einstein's theory gives exactly the same result as the Newton's
theory whereas in the Newtonian case such second-order equations are
in fact exactly valid to fully nonlinear order. Therefore, any
nonvanishing third and higher order perturbations in Einstein's
gravity are supposed to be the pure relativistic corrections. We
will present the scalar-type perturbation equations to the third
order and the gravitational wave equation to the second order. This
is enough because the previously known gravitational wave equation
will already get the correction terms even in the second order,
whereas for the scalar-type perturbation new correction terms appear
only in the third order compared with the previous studies
\cite{second-order}. If we include the gravitational waves, the
gravitational waves only to the second order will be needed to make
the third-order scalar type perturbation equations complete.

We will {\it ignore} the vector-type perturbation in this work, thus
consider only irrotational perturbation, because due to the angular
momentm conservation the rotational perturbation always decays in
expanding phase; this situation will be reversed in the collapsing
phase where the rotational mode will grow again due to the angular
momentm conservation. We will consider a {\it flat} background with
two reasons. {}First, even in the linear perturbations it is known
that the relativistic-Newtonian correspondence is rather ambiguous
in the presence of the background curvature; this is the case if we
include the gravitational potential in making the correspondence
\cite{HN-Newtonian-1999}. Second, current observations of the
large-scale structure and the cosmic microwave background radiation
favour near flat Friedmann world model with non-vanishing
cosmological constant \cite{observations}. In fact, we will {\it
include} the cosmological constant in our analyses, thus our results
are relevant to currently favoured cosmology.

Our previous study on the second-order perturbations revealed which
gauge condition (equivalently, gauge-invariant combinations) suits
our problem \cite{second-order}. In fact, unique gauge conditions
were distinguished in showing the correspondence: these were the
spatial $C$-gauge and the temporal comoving gauge. The spatial
$C$-gauge was a rather natural choice because only in this gauge the
spatial gauge mode can be fixed completely to {\it all} orders. In
this gauge all the rest of the variables can be equivalently
regarded as spatially gauge-invariant ones to all orders, see
\cite{NL}. In the temporal gauge choice, however, we have many (in
fact, infinitely many) different gauge choices, which fix the
temporal gauge mode completely. In any of such gauge conditions all
the rest of the variables can be equivalently regarded as temporally
gauge-invariant ones. Such procedures for taking the temporal gauge
condition are separately available to each order in perturbations;
i.e., we can choose different gauge conditions to different
perturbational order, see \cite{NL}.

The (temporal) comoving gauge distinguished itself with the
following reasons. In \cite{second-order} we have successfully shown
that in the comoving gauge we can identify the density and velocity
variables which allow us to derive relativistic equations identical
to the Newtonian ones. This point, perhaps, does not necessarily
imply that it is not possible to discover any other variables and
gauges which also lead to the same identification. But, we do mean
that the comoving gauge is a natural choice. This also does not
necessarily imply that even to the third order the comoving gauge
choice will be the best choice. As a matter of fact, our
policy/strategy about the gauge choice is that we do not know which
gauge will suit the problem before we investigate and try many
different gauge conditions. That is why we have presented our basic
set of equations in a gauge-ready form which allows maximal usage of
many different gauge conditions: see \cite{Bardeen-1988,H-1991} for
the linear case and \cite{NL} for the second-order perturbations.

In this work, based on our successful experiences in the linear and
second-order perturbations we will take the temporal comoving gauge
condition and the spatial $C$-gauge even to the third order. First,
under this gauge condition we can write the fully nonlinear
equations in simple forms. Second, only in this gauge we have the
relativistic equations the same as the Newtonian ones to the second
order. And third, quite interestingly, we will show that in order to
derive the third-order perturbations in this gauge, in fact, we need
to evaluate the geometric and energy-momentum variables to the
second order only. This is possible due to our appropriate choice of
the variables and the gauge conditions. Thus, in the following we
will derive the third-order pure relativistic correction terms in
the comoving gauge condition. As we mentioned, since our gauge
conditions fix the gauge modes completely, each of the variables we
are using has a unique corresponding gauge-invariant combination.
Thus, our analyses can be equivalently regarded as gauge-invariant
ones extended to the third-order perturbations.

Here, we summarize the relativistic-Newtonian correspondences known up to
second-order perturbations in a flat Friedmann world model without pressure.
To the background order we have \cite{Friedmann-1922,Milne-1934}
\bea
   & & H^2 = {8 \pi G \over 3} \mu
       - {{\rm const.} \over a^2} + {\Lambda \over 3},
   \label{BG}
\eea with the energy (mass) density $\mu$ ($\varrho$) $\propto
a^{-3}$; $a(t)$ is the cosmic scale factor and $H \equiv {\dot a /
a}$; we set $c \equiv 1$. The ``{\rm const.}'' part is interpreted
as the spatial curvature in Einstein's gravity
\cite{Friedmann-1922}, and the total energy in the Newton's gravity
\cite{Milne-1934}. We put the cosmological constant $\Lambda$ by
hand which can work as a repulsive force proportional to the
distance for $\Lambda > 0$. To the linear-order perturbations we
have \cite{Lifshitz-1946,Bonnor-1957} \bea
   & & \ddot \delta + 2 H \dot \delta - 4 \pi G \varrho \delta = 0,
   \label{pert-1st}
\eea where $\delta \equiv {\delta \mu / \mu} = {\delta \varrho /
\varrho}$ with $\mu$ ($\varrho$) and $\delta \mu$ ($\delta \varrho$)
the background and perturbed parts of the energy (mass) density.
This equation is valid considering general presence of the
background spatial curvature term and the cosmological constant.
Now, to the second order, assuming flat background, we have
\cite{Peebles-1980,NL,second-order} \bea
   \dot \delta + {1 \over a} \nabla \cdot {\bf u}
   &=& - {1 \over a} \nabla \cdot \left( \delta {\bf u} \right),
   \label{dot-delta-eq-2nd} \\
   \nabla \cdot \left( \dot {\bf u} + H {\bf u} \right)
       + 4 \pi G \varrho a \delta
   &=& - {1 \over a} \nabla \cdot \left( {\bf u} \cdot \nabla {\bf u} \right),
   \label{dot-u-eq-2nd}
\eea
or by combining these, we have
\bea
   & & \ddot \delta
       + 2 H \dot \delta - 4 \pi G \varrho \delta
   \nonumber \\
   & & \qquad
       = - {1 \over a^2} {\partial \over \partial t}
       \left[ a \nabla \cdot \left( \delta {\bf u} \right) \right]
       + {1 \over a^2} \nabla \cdot \left( {\bf u} \cdot
       \nabla {\bf u} \right).
   \label{ddot-delta-eq-2nd}
\eea
These equations are valid in the presence of cosmological constant.
The above equations are valid in both Einstein's and Newton's theories.
In the relativistic theory we have to specify the variables
$\delta$ and ${\bf u}$ which correspond to the relative density fluctuation
and perturbed velocity in Newton's theory; this will be done later.

We believe no one would have anticipated such an exact coincidence
to the second-order perturbations especially considering the
presence of the horizon in the relativistic treatment. It might
happen as well that our relativistic results give relativistic
correction terms appearing to the second order which could become
important as the scale approaches and goes beyond the horizon. Our
results show that there are {\it no} such correction terms appearing
to the second order, and the correspondence is {\it exact} to that
order. Equations (\ref{dot-delta-eq-2nd})-(\ref{ddot-delta-eq-2nd})
are valid in fully nonlinear situation in Newton's theory
\cite{Peebles-1980} whereas these are valid only up to the second
order in Einstein's case \cite{NL,second-order}. It is our task to
derive the third-order correction terms in Einstein's theory and to
show how it causes difference between the two theories.

%
%
\section{Fully nonlinear equations}
                                                \label{sec:NL}

In \cite{second-order} we have presented the fully nonlinear
equations in the comoving gauge condition using the $1+3$ covariant
formulation \cite{covariant} and the $3+1$ ADM
(Arnowitt-Deser-Misner) \cite{ADM} formulations of Einstein's
gravity. In \cite{second-order} we showed that these two
formulations are equivalent, and in the following we will take the
ADM formulation. The basic set of ADM equations in our notation can
be found in \S II.A of \cite{NL}, see also \cite{Bardeen-1980}. As
this work can be regarded as a continuation of our previous studies
in \cite{NL,second-order}, we follow the notations used in that
works.

In the ADM approach, the temporal comoving gauge condition to all
orders sets the flux four-vector to vanish, i.e., $J_\alpha \equiv
0$; here we also used the irrotational condition which ignores the
vector-type perturbation. In \cite{NL} the fluid quantities are
introduced based on the normal frame four-vector $\tilde n_a$ with
$\tilde n_\alpha \equiv 0$; in this case the information of the
fluid motion is present in the flux vector $\tilde q_a$ with $\tilde
q_a \tilde n^a \equiv 0$. In such a choice of the frame the temporal
comoving gauge condition with vanishing rotation implies $\tilde
u_\alpha \equiv 0$ of the fluid four-vector $\tilde u_a$. Thus, the
fluid four-vector {\it coincides} with the normal four-vector. The
physical zero-pressure condition implies vanishing isotropic
pressure ($\tilde p$ or $S$ in the ADM notation) and anisotropic
stress ($\tilde \pi_{ab}$ or $\bar S_{\alpha\beta}$), i.e., $\tilde
p \equiv 0 \equiv \tilde \pi_{ab}$ or $S \equiv 0 \equiv \bar
S_{\alpha\beta}$ to all orders, based on the fluid four-vector. As
the $\tilde u_a$ coincides with the $\tilde n_a$ in our comoving
gauge, the zero-pressure conditions simply allow us to set all the
pressure terms in the normal-frame fluid quantites equal to zero; we
point out that this is not true in the other gauge conditions in the
normal-frame fluid quantities, see \cite{second-order}; in the
normal frame, in other than the comoving gauge conditions the
pressure and anisotropic stress do not vanish to the second and
higher orders even in the physically zero-pressure situation; for
further discussions, see \cite{second-order}. Although we do not
need the form of energy-momentum tensor, from Eqs.\ (3), (4) of
\cite{NL} we have \bea
   \tilde T^{00} = N^{-2} E, \quad
       \tilde T^0_\alpha = 0, \quad
       \tilde T_{\alpha\beta} = 0,
   \label{Tab}
\eea where tildes indicate the covariant quantities; $E$ is the ADM
energy density and $N$ is the lapse function defined as $N^2 \equiv
- (\tilde g^{00})^{-1}$.

The momentum conservation equation in Eq.\ (13) of \cite{NL} gives
\bea
   & & N_{,\alpha} = 0.
\eea Thus, we may set $N \equiv a(t)$ to all orders. In this case we
have $\dot E \equiv E_{,0} N^{-1}$. The energy conservation equation
and the trace part of ADM propagation equation in Eqs.\ (12), (10)
of \cite{NL} give \bea
   \hat {\dot E} - K E
   &=& 0,
   \label{ADM-eq1} \\
   \hat {\dot K} - { 1\over 3} K^2 - \bar K^{\alpha\beta} \bar K_{\alpha\beta}
       - 4 \pi G E + \Lambda
   &=& 0,
   \label{ADM-eq2}
\eea
where $\hat {\dot E} \equiv \dot E - E_{,\alpha} N^\alpha N^{-1}$, {\it etc.};
$K$ and $\bar K_{\alpha\beta}$ are the trace and tracefree parts, respectively,
of the extrinsic curvature $K_{\alpha\beta}$ of the normal hypersurface
introduced in the ADM formulation, and $N_\alpha$ is a shift vector defined as
$N_\alpha \equiv \tilde g_{0\alpha}$.
The spatial indices in the ADM formulation are based on the ADM
three-space metric $h_{\alpha\beta}$ defined as
$h_{\alpha\beta} \equiv \tilde g_{\alpha\beta}$.
By combining these equations we have
\bea
   & & \left( {\hat {\dot E} \over E} \right)^{\hat \cdot}
       - {1 \over 3} \left( {\hat {\dot E} \over E} \right)^2
       - \bar K^{\alpha\beta} \bar K_{\alpha\beta}
       - 4 \pi G E + \Lambda
       = 0.
   \label{ADM-eq3}
\eea Equations (\ref{Tab})-(\ref{ADM-eq3}) are valid to all orders,
i.e., these equations are fully nonlinear.

%
%
\section{Third order perturbations}
                                                \label{sec:Third}

We consider the scalar- and tensor-type perturbations in the flat Friedmann
background.
As the metric we take
\bea
   ds^2
   &=& - a^2 \left( 1 + 2 \alpha \right) d \eta^2
       - 2 a^2 \beta_{,\alpha} d \eta d x^\alpha
   \nonumber \\
   & &
       + a^2 \left[ g^{(3)}_{\alpha\beta} \left( 1 + 2 \varphi \right)
       + 2 \gamma_{,\alpha|\beta}
       + 2 C^{(t)}_{\alpha\beta} \right] d x^\alpha d x^\beta,
   \label{metric}
\eea where $\alpha$, $\beta$, $\gamma$ and $\varphi$ are spacetime
dependent perturbed-order variables; we take Bardeen's metric
convention in \cite{Bardeen-1988,H-1991} extended to the third
order. A vertical bar indicates a covariant derivative based on
$g^{(3)}_{\alpha\beta}$ which can be regarded as
$\delta_{\alpha\beta}$ if we use Cartesian coordinates in the flat
Friedmann background. By taking $\gamma \equiv 0$, which we call the
spatial $C$-gauge, the spatial gauge mode is removed completely,
thus all the remaining variables we are using are spatially
gauge-invariant to the third order; this is true if we
simultaneously take a temporal gauge which removes the temporal
gauge mode completely, see \S VI.B.2 and C.1 of \cite{NL}. In the
following we will take $\gamma \equiv 0$ as the spatial gauge
condition and use $\chi \equiv a \beta + a^2 \dot \gamma$ which
becomes $\chi = a \beta$.

We expand
\bea
   & & E \equiv \mu + \delta \mu, \quad
       K \equiv - 3 H + \kappa.
   \label{E-K}
\eea
Up to this point our notations look exactly the same as in the linear theory
whereas, in fact, we consider each perturbation variable to be expanded to the
third order.
As an example, for $\delta \mu$ we have
\bea
   & & \delta \mu \equiv \delta \mu^{(1)} + \delta \mu^{(2)}
       + \delta \mu^{(3)} + \dots,
\eea where, to the third order we truncate the expansion at
third-order term $\delta \mu^{(3)}$. A close examination of our
fully nonlinear equations in \S \ref{sec:NL} reveals an important
technical methods of performing the third-order perturbations by
using only the second-order expansion of perturbation variables.
Such a simple method is possible due to our right choice of the
equations, the right gauge conditions, and our proper choice of the
fundamental variables to be matched with the Newtonian variables to
the third order. That is, in our calculations we do not even need to
have the inverse metric expanded to the third order, and all
algebraic quantities we need can be found in \cite{NL} which
presents various useful quantities expanded to the second order in
perturbations.

To the {\it linear order} we identified \cite{HN-Newtonian-1999}
\bea
   & & \delta \varrho \equiv \delta \mu_v, \quad
       \delta \Phi \equiv - \varphi_\chi = \alpha_\chi,
   \nonumber \\
   & & {\bf u} \equiv - \nabla v_\chi, \quad
       - {1 \over a} \nabla \cdot {\bf u}
       \equiv {\Delta \over a} v_\chi
       \equiv \kappa_v.
   \label{identify-linear-order}
\eea To the {\it second order} we identified \cite{second-order}
\bea
   & & \delta \mu_v \equiv \delta \varrho, \quad
       \kappa_v \equiv - {1 \over a} \nabla \cdot {\bf u}.
   \label{identify-second-order}
\eea Based on our experience in the second-order perturbations, we
have in mind to identify Eq.\ (\ref{identify-second-order}) even to
the {\it third order}. It may turn out to be that these are not the
best identifications, but in the following we will {\it assume}
these are the right ones and will {\it take} the consequent
additional third-order terms as the pure relativistic corrections.

The perturbed-order variable $v$ is defined as $J_\alpha \equiv - a
\mu v_{,\alpha}$ to all orders, thus $v$ is related to the velocity
or flux variable. Our comoving gauge condition sets $v \equiv 0$. In
our notation, $\delta \mu_v$ indicates a gauge-invariant combination
which is the same as $\delta \mu$ in the comoving gauge which sets
$v \equiv 0$. Such a variable (equivalently, a gauge-invariant
combination) is unique to all orders in perturbations; to the linear
order we have $\delta \mu_v \equiv \delta \mu - \dot \mu a v$, and
to the second order see Eq.\ (282) in \cite{NL}. Similarly,
$\varphi_\chi$ is a gauge-invariant combination equivalent to
$\varphi$ in the zero-shear gauge which sets $\chi \equiv 0$; to the
linear order we have $\varphi_\chi \equiv \varphi - H \chi$, and to
the second order, see Eq.\ (280) in \cite{NL}. {}For our
justification to name the gauge conditions, see below Eq.\ (265) of
\cite{NL}.

One of the terms we need to evaluate to the third order in Eqs.\
(\ref{ADM-eq1})-(\ref{ADM-eq3}) is $\bar K^{\alpha\beta} \bar
K_{\alpha\beta}$. Since $\bar K_{\alpha\beta}$ is already at least
linear order, we need to evaluate $\bar K^{\alpha\beta}$ to the
second order only, etc. The other ones we need to evaluate to the
third order are \bea
   & & \hat {\dot E} \equiv \dot E - E_{,\alpha} N^\alpha N^{-1}, \quad
       \hat {\dot K} \equiv \dot K - K_{,\alpha} N^\alpha N^{-1}.
\eea Due to our identifications in Eq.\
(\ref{identify-second-order}), we are using the perturbed parts of
$E$ and $K$ in Eq.\ (\ref{E-K}) as the fundamental perturbation
variables. In order to evaluate $E_{,\alpha} N^\alpha$, since
$E_{,\alpha}$ (or $N^\alpha$) is already at least linear order, it
is enough to evaluate $N^\alpha$ (or $E_{,\alpha}$) only to the
second order, and similarly for $K_{,\alpha} N^\alpha$. Thus, using
Eqs.\ (55), (57), and (175) of \cite{NL}, we have \bea
   & & E_{,\alpha} N^\alpha
       = - {1 \over a} \delta \mu_{,\alpha} \chi^{,\alpha}
       \left( 1 - 2 \varphi \right)
       + 2 {1 \over a} \delta \mu^{,\alpha} \chi^{,\beta} C^{(t)}_{\alpha\beta},
   \\
   & & K_{,\alpha} N^\alpha
       = - {1 \over a} \kappa_{,\alpha} \chi^{,\alpha}
       \left( 1 - 2 \varphi \right)
       + 2 {1 \over a} \kappa^{,\alpha} \chi^{,\beta} C^{(t)}_{\alpha\beta},
   \\
   & & \bar K^{\alpha\beta} \bar K_{\alpha\beta}
       = \left[ {1 \over a^2} \left( \chi^{,\alpha|\beta}
       - {1 \over 3} g^{(3)\alpha\beta} \Delta \chi \right)
       + \dot C^{(t)\alpha\beta} \right]
   \nonumber \\
   & & \qquad
       \times
       \Bigg\{ \left( {1 \over a^2} \chi_{,\alpha|\beta}
       + \dot C^{(t)}_{\alpha\beta} \right)
       \left( 1 - 2 \alpha - 4 \varphi \right)
       - {4 \over a^2} \chi_{,\alpha} \varphi_{,\beta}
   \nonumber \\
   & & \qquad
       - 4 \dot \varphi C^{(t)}_{\alpha\beta}
       - {2 \over a^2} \chi^{,\gamma} \left(
       2 C^{(t)}_{\gamma\alpha|\beta}
       - C^{(t)}_{\alpha\beta|\gamma} \right)
   \nonumber \\
   & & \qquad
       - 4 C^{(t)\gamma} _{\;\;\;\;\;\alpha}
       \left( {1 \over a^2} \chi_{,\beta|\gamma}
       + \dot C^{(t)}_{\beta\gamma} \right) \Bigg\}.
   \label{shear-square}
\eea Thus, Eqs.\ (\ref{ADM-eq1}), (\ref{ADM-eq2}) give \bea
   & & \left( {\dot \mu \over \mu} + 3 H \right)
       \left( 1 + \delta \right)
       + \dot \delta
       - \kappa
   \nonumber \\
   & & \qquad
       = \kappa \delta
       - {1 \over a^2} \delta_{,\alpha} \chi^{,\alpha}
       \left( 1 - 2 \varphi \right)
       + 2 {1 \over a^2} \delta^{,\alpha} \chi^{,\beta}
       C^{(t)}_{\alpha\beta},
   \label{dot-delta-eq} \\
   & & - 3 \dot H - 3 H^2
       - 4 \pi G \mu + \Lambda
       + \dot \kappa + 2 H \kappa - 4 \pi G \delta \mu
   \nonumber \\
   & & \qquad
       = {1 \over 3} \kappa^2
       - {1 \over a^2} \kappa_{,\alpha} \chi^{,\alpha}
       \left( 1 - 2 \varphi \right)
       + 2 {1 \over a^2} \kappa^{,\alpha} \chi^{,\beta} C^{(t)}_{\alpha\beta}
   \nonumber \\
   & & \qquad
       + \left[ {1 \over a^2} \left( \chi^{,\alpha|\beta}
       - {1 \over 3} g^{(3)\alpha\beta} \Delta \chi \right)
       + \dot C^{(t)\alpha\beta} \right]
   \nonumber \\
   & & \qquad
       \times
       \Bigg[ \left( {1 \over a^2} \chi_{,\alpha|\beta}
       + \dot C^{(t)}_{\alpha\beta} \right)
       \left( 1 - 2 \alpha - 4 \varphi \right)
   \nonumber \\
   & & \qquad
       - {4 \over a^2} \chi_{,\alpha} \varphi_{,\beta}
       - 4 \dot \varphi C^{(t)}_{\alpha\beta}
       - {2 \over a^2} \chi^{,\gamma} \left(
       2 C^{(t)}_{\gamma\alpha|\beta}
       - C^{(t)}_{\alpha\beta|\gamma} \right)
   \nonumber \\
   & & \qquad
       - 4 C^{(t)\gamma} _{\;\;\;\;\;\alpha}
       \left( {1 \over a^2} \chi_{,\beta|\gamma}
       + \dot C^{(t)}_{\beta\gamma} \right) \Bigg].
   \label{dot-kappa-eq}
\eea We note that $\alpha$ and $\dot \varphi$ in the comoving gauge
are already quadratic order at least, thus can be ignored in Eq.\
(\ref{dot-kappa-eq}); see Eqs.\ (12), (20) of \cite{second-order}.

Now, we need Newtonian expressions of $\varphi$ to the linear order
and $\chi$ to the second order.
Since we are considering the comoving gauge condition our $\varphi$ and $\chi$
are the same as the gauge-invariant combinations $\varphi_v$ and $\chi_v$,
respectively.
To the linear order we have
\bea
   & & \varphi_v \equiv \varphi - a H v
       = \varphi_\chi - a H v_\chi,
   \nonumber \\
   & & \chi_v \equiv \chi - a v \equiv - a v_\chi.
   \label{GI}
\eea To the second order, $\chi_v$ is presented in Eq.\ (284) of
\cite{NL}. To the second order, from Eq.\ (197) of \cite{NL} we have
\bea
   & & \kappa + {\Delta \over a^2} \chi
       = N^{(s)}_{2} \Big|_v
       = {1 \over a^2} \left( 2 \varphi \Delta \chi
       - \chi^{,\alpha} \varphi_{,\alpha} \right)
   \nonumber \\
   & & \qquad
       + {3 \over 2} {1 \over a^2} \Delta^{-1} \nabla^\alpha \left(
       \chi_{,\alpha} \Delta \varphi
       + \chi^{,\beta} \varphi_{,\alpha|\beta} \right)
       + {2 \over a^2} \chi^{,\beta|\gamma} C^{(t)}_{\beta\gamma}
   \nonumber \\
   & & \qquad
       + {3 \over 2} \Delta^{-1} \nabla^\alpha \Big(
       - \varphi^{,\beta} \dot C^{(t)}_{\alpha\beta}
       + {1 \over a^2} \chi^{,\beta} \Delta C^{(t)}_{\alpha\beta}
   \nonumber \\
   & & \qquad
       + 2 C^{(t)\beta\gamma} \dot C^{(t)}_{\alpha\beta|\gamma}
       - C^{(t)\beta\gamma}
       \dot C^{(t)}_{\beta\gamma|\alpha} \Big)
       + {1 \over 2} C^{(t)\beta\gamma} \dot C^{(t)}_{\beta\gamma}
   \nonumber \\
   & & \qquad
       \equiv {1 \over a} X,
   \label{kappa-chi-eq}
\eea
where we have ignored $\alpha$ and $\dot \varphi$ which
contribute to the third order.
Apparently, we also need $\dot C^{(t)}_{\alpha\beta}$ to the second order.
This will be presented in \S \ref{sec:GW}.

%
%
\section{Scalar-type corrections}
                                                \label{sec:Scalar}

Ignoring $C^{(t)}_{\alpha\beta}$, the perturbed parts of Eqs.\
(\ref{dot-delta-eq}), (\ref{dot-kappa-eq}), and (\ref{kappa-chi-eq})
give the complete set. As we ignore the rotational type perturbation
the Newtonian velocity perturbation ${\bf u}$ is of a potential
type, i.e., ${\bf u} = \nabla u$. Thus, Eqs.\
(\ref{identify-second-order}), (\ref{kappa-chi-eq}) give \bea
   & & \kappa \equiv - {1 \over a} \nabla \cdot {\bf u}
       = - {1 \over a} \Delta u, \quad
       \chi = a u + a \Delta^{-1} X,
   \label{identify-chi}
\eea where $\kappa$ is valid to the third order and $\chi$ is valid
to the second order. Thus, Eqs.\ (\ref{dot-delta-eq}),
(\ref{dot-kappa-eq}), and (\ref{kappa-chi-eq}) can be written as
\bea
   & & \dot \delta + {1 \over a} \nabla \cdot {\bf u}
       = - {1 \over a} \nabla \cdot \left( \delta {\bf u} \right)
   \nonumber \\
   & & \quad
       + {1 \over a} \left[ 2 \varphi {\bf u}
       - \nabla \left( \Delta^{-1} X \right) \right] \cdot \nabla \delta,
   \label{delta-eq-3rd} \\
   & & {1 \over a} \nabla \cdot \left( \dot {\bf u}
       + H {\bf u} \right)
       + 4 \pi G \mu \delta
       = - {1 \over a^2} \nabla \cdot \left( {\bf u}
       \cdot \nabla {\bf u} \right)
   \nonumber \\
   & & \quad
       - {2 \over 3 a^2} \varphi
       {\bf u} \cdot \nabla \left( \nabla \cdot {\bf u} \right)
       + {4 \over a^2} \nabla \cdot \left[ \varphi
       \left( {\bf u} \cdot \nabla {\bf u}
       - {1 \over 3} {\bf u} \nabla \cdot {\bf u} \right) \right]
   \nonumber \\
   & & \quad
       - {\Delta \over a^2}
       \left[ {\bf u} \cdot \nabla \left( \Delta^{-1} X \right) \right]
       + {1 \over a^2} {\bf u} \cdot \nabla X
       + {2 \over 3a^2} X \nabla \cdot {\bf u},
   \label{u-eq-3rd}
\eea
where
\bea
   X
   &\equiv&
       2 \varphi \nabla \cdot {\bf u}
       - {\bf u} \cdot \nabla \varphi
       + {3 \over 2} \Delta^{-1} \nabla \cdot
       \left[ {\bf u} \cdot \nabla \left( \nabla \varphi \right)
       + {\bf u} \Delta \varphi \right].
   \nonumber \\
   \label{X-eq-3rd}
\eea Equations (\ref{delta-eq-3rd}), (\ref{u-eq-3rd}) extend Eqs.\
(\ref{dot-delta-eq-2nd}), (\ref{dot-u-eq-2nd}) to the third order.
By combining Eqs.\ (\ref{delta-eq-3rd}), (\ref{u-eq-3rd}) we can
derive \bea
   & & \ddot \delta + 2 {\dot a \over a} \dot \delta
       - 4 \pi G \mu \delta
       = - {1 \over a^2} {\partial \over \partial t}
       \left[ a \nabla \cdot \left( \delta {\bf u} \right) \right]
       + {1 \over a^2} \nabla \cdot \left( {\bf u}
       \cdot \nabla {\bf u} \right)
   \nonumber \\
   & & \quad
       + {1 \over a^2} {\partial \over \partial t}
       \left\{ a \left[ 2 \varphi {\bf u}
       - \nabla \left( \Delta^{-1} X \right) \right] \cdot \nabla \delta
       \right\}
   \nonumber \\
   & & \quad
       + {2 \over 3 a^2} \varphi
       {\bf u} \cdot \nabla \left( \nabla \cdot {\bf u} \right)
       - {4 \over a^2} \nabla \cdot \left[ \varphi
       \left( {\bf u} \cdot \nabla {\bf u}
       - {1 \over 3} {\bf u} \nabla \cdot {\bf u} \right) \right]
   \nonumber \\
   & & \quad
       + {\Delta \over a^2}
       \left[ {\bf u} \cdot \nabla \left( \Delta^{-1} X \right) \right]
       - {1 \over a^2} {\bf u} \cdot \nabla X
       - {2 \over 3a^2} X \nabla \cdot {\bf u},
   \label{density-eq-3rd}
\eea which extends Eq.\ (\ref{ddot-delta-eq-2nd}) to the third
order. The last three lines of Eq.\ (\ref{density-eq-3rd}) are the
third-order terms. Examination of Eqs.\
(\ref{delta-eq-3rd})-(\ref{X-eq-3rd}) shows that all the third-order
correction terms are of $\varphi$-order higher than the second-order
terms like ${1 \over a} \nabla \cdot \left( \delta {\bf u} \right)$
and ${1 \over a^2} \nabla \cdot \left( {\bf u} \cdot \nabla {\bf u}
\right)$. As the fully nonlinear zero-pressure Newtonian equations
are exact to the second order, the above third-order correction
terms in our relativistic analyses are pure relativistic correction
terms. Therefore, the pure general relativistic effects are at least
$\varphi$-order higher than the Newtonian ones. Up to the
third-order corrections appearing in the general relativity, the
effects are {\it independent} of the horizon scale and depend on the
linear-order curvature (gravitational potential, see below)
perturbation strength {\it only}.

In Eqs.\ (\ref{delta-eq-3rd})-(\ref{X-eq-3rd}) we need $\varphi$
only to the linear order. Thus, let us examine the behavior of
$\varphi$ to the linear order. In our comoving gauge condition
$\varphi$ is equivalent to a gauge-invariant combination
$\varphi_v$, and to the linear order from Eq.\ (\ref{GI}) we have
\bea
   & & \varphi_v = \varphi_\chi - a H v_\chi,
   \label{varphi_v}
\eea where we have $\varphi_\chi = - \delta \Phi$ and ${\bf u} = -
\nabla v_\chi$ in Eq.\ (\ref{identify-linear-order}). Thus, in terms
of the Newtonian variables we have \bea
   & & \varphi = - \delta \Phi
       + \dot a \Delta^{-1} \nabla \cdot {\bf u}.
\eea Exact solutions of linear perturbation were presented in Tables
of \cite{H-MDE-1994}. {}For $K = 0$ we have \cite{H-MDE-1994} \bea
   & & \varphi_v = C,
   \nonumber \\
   & & \varphi_\chi = - \alpha_\chi
       = - 4 \pi G \mu a^2 \Delta^{-1} \delta_v
       = C 4 \pi G \mu a^2 H \int^t {dt \over \dot a^2},
   \nonumber \\
   & & v_\chi = a \Delta^{-1} \kappa_v
       = - C {1 \over aH}
       \left( 1 + a^2 H \dot H \int^t {dt \over \dot a^2} \right),
\eea where the lower bounds of integrations give decaying modes.
{}For $K = 0 = \Lambda$, using $a \equiv a_1 t^{2/3}$, we have
\cite{H-MDE-1994} \bea
   & & \varphi_v = C,
   \nonumber \\
   & & \varphi_\chi = - \alpha_\chi
       = - {2 \over 3} a_1^2 \Delta^{-1} \delta_v
       = {3 \over 5} C + {4 \over 9} d t^{-5/3},
   \nonumber \\
   & & v_\chi = a_1 t^{2/3} \Delta^{-1} \kappa_v
       = - {1 \over a_1}
       \left( {3 \over 5} C t^{1/3} - {2 \over 3} d t^{-4/3} \right).
\eea Notice that to the linear order we have \bea
   & & \varphi_v = C,
\eea and $\varphi_v$ has {\it no} decaying mode in expanding phase;
this is true considering the presence of the cosmological constant.
In fact, to the linear order $\varphi_v$ satisfies
\cite{HN-Newtonian-1999} \bea
   & & \dot \varphi_v = 0.
\eea See also Eq.\ (20) in the second reference of
\cite{second-order}. Ignoring the decaying mode, for $\Lambda = 0$,
we have \bea
   & & \varphi_\chi = {3 \over 5} \varphi_v,
\eea
and the temperature anisotropy of cosmic microwave background radiation (CMB)
gives \cite{SW-1967,HN-SW1}
\bea
   & & {\delta T \over T} \sim {1 \over 3} \varphi_\chi
       \sim {1 \over 5} \varphi_v \sim {1 \over 5} C,
   \label{SW}
\eea
to the linear order.
This is a part of the Sachs-Wolfe effect in a flat background without
the cosmological constant.
The observations of CMB give ${\delta T / T} \sim 10^{-5}$ \cite{CMB}, thus
\bea
   & & \varphi_v \sim 5 \times 10^{-5},
   \label{CMB-constraint}
\eea
in the large-scale limit near horizon scale.
Our $\varphi$ is dimensionless.

We call $\varphi$ the curvature perturbation because it is related
to the perturbed part of the spatial curvature of the normal hypersurface.
To the linear order we have
\bea
   & & R^{(h)} = {6 \bar K \over a^2}
       - 4 {\Delta + 3 \bar K \over a^2} \varphi,
\eea where $R^{(h)}$ is the scalar-curvature of the three-space
metric $h_{\alpha\beta}$, and $\bar K$ is the sign of the background
curvature, see Eq.\ (4) in \cite{H-1991}. To the second order, see
Eq.\ (265) in \cite{NL}. In fact, from Eqs.\ (7), (55), and (175) of
\cite{NL} we can easily check that $\varphi$ characterizes the
three-space Riemann curvature
$R^{(h)\alpha}_{\;\;\;\;\;\;\;\beta\gamma\delta}$ to {\it all
orders} in perturbations; this is true assuming pure scalar-type
perturbation. The gauge-invariant combination $\varphi_v$ is also
known to be one of the large-scale (super-sound-horizon) limit
conserved variables in various situations including time varying
equation of state, field potential, and generalized gravity theories
\cite{conserved}. It is also known to be conserved in the
large-scale (super-sound-horizon) limit even in nonlinear situations
\cite{Salopek-Bond-1990,NL}.

%
%
\section{Including the gravitational waves}
                                                  \label{sec:GW}

In this section we present the complete set of equations to the
third order, now including the contribution of gravitational waves,
and also the gravitational wave equation complete to the second
order. Using Eqs.\ (\ref{identify-linear-order}),
(\ref{identify-second-order}), and (\ref{identify-chi}), Eqs.\
(\ref{dot-delta-eq}), (\ref{dot-kappa-eq}), and (\ref{kappa-chi-eq})
give \bea
   & & \dot \delta + {1 \over a} \nabla \cdot {\bf u}
       = - {1 \over a} \nabla \cdot \left( \delta {\bf u} \right)
   \nonumber \\
   & & \quad
       + {1 \over a} \left[ 2 \varphi {\bf u}
       - \nabla \left( \Delta^{-1} X \right) \right] \cdot \nabla \delta
       + {2 \over a} \delta^{,\alpha} u^\beta C^{(t)}_{\alpha\beta},
   \label{delta-eq-3rd-GW} \\
   & & {1 \over a} \nabla \cdot \left( \dot {\bf u} + H {\bf u} \right)
       + 4 \pi G \mu \delta
   \nonumber \\
   & & \quad
       = - {1 \over a^2} \nabla \cdot \left( {\bf u}
       \cdot \nabla {\bf u} \right)
       - \dot C^{(t)\alpha\beta} \left( {2 \over a} u_{\alpha|\beta}
       + \dot C^{(t)}_{\alpha\beta} \right)
   \nonumber \\
   & & \quad
       - {2 \over 3 a^2} \varphi
       {\bf u} \cdot \nabla \left( \nabla \cdot {\bf u} \right)
       + {4 \over a^2} \nabla \cdot \left[ \varphi
       \left( {\bf u} \cdot \nabla {\bf u}
       - {1 \over 3} {\bf u} \nabla \cdot {\bf u} \right) \right]
   \nonumber \\
   & & \quad
       - {\Delta \over a^2}
       \left[ {\bf u} \cdot \nabla \left( \Delta^{-1} X \right) \right]
       + {1 \over a^2} {\bf u} \cdot \nabla X
       + {2 \over 3a^2} X \nabla \cdot {\bf u}
   \nonumber \\
   & & \quad
       + {2 \over a^2} \left( \nabla \cdot {\bf u} \right)^{,\alpha} u^\beta
       C^{(t)}_{\alpha\beta}
       + {4 \over a} \dot C^{(t)\alpha\beta}
   \nonumber \\
   & & \quad
       \times
       \left[ u_\alpha \varphi_{,\beta}
       + 2 \varphi u_{\alpha|\beta}
       - {1 \over 2} \left( \Delta^{-1} X \right)_{,\alpha|\beta}
       + a \varphi \dot C^{(t)}_{\alpha\beta} \right]
   \nonumber \\
   & & \quad
       + 4 \left( {1 \over a} u^{\alpha|\beta}
       + \dot C^{(t)\alpha\beta} \right)
       \Bigg[ C^{(t)\gamma}_{\;\;\;\;\;\alpha}
       \left( {1 \over a} u_{\beta|\gamma}
       + \dot C^{(t)}_{\beta\gamma} \right)
   \nonumber \\
   & & \quad
       - {1 \over 3a} C^{(t)}_{\alpha\beta} \nabla \cdot {\bf u}
       + {1 \over 2 a} \left( 2 C^{(t)}_{\gamma\alpha|\beta}
       - C^{(t)}_{\alpha\beta|\gamma} \right) u^\gamma
       \Bigg],
   \label{u-eq-3rd-GW}
\eea
where
\bea
   & & X \equiv 2 \varphi \nabla \cdot {\bf u}
       - {\bf u} \cdot \nabla \varphi
       + {3 \over 2} \Delta^{-1} \nabla \cdot
       \left[ {\bf u} \cdot \nabla \left( \nabla \varphi \right)
       + {\bf u} \Delta \varphi \right]
   \nonumber \\
   & & \qquad
       + 2 u^{\beta|\gamma} C^{(t)}_{\beta\gamma}
       + {3 \over 2} a \Delta^{-1} \nabla^\alpha \Big(
       - \varphi^{,\beta} \dot C^{(t)}_{\alpha\beta}
       + {1 \over a} u^{\beta} \Delta C^{(t)}_{\alpha\beta}
   \nonumber \\
   & & \qquad
       + 2 C^{(t)\beta\gamma} \dot C^{(t)}_{\alpha\beta|\gamma}
       - C^{(t)\beta\gamma}
       \dot C^{(t)}_{\beta\gamma|\alpha} \Big)
   \nonumber \\
   & & \qquad
       + {1 \over 2} a C^{(t)\beta\gamma} \dot C^{(t)}_{\beta\gamma}.
   \label{X-eq-3rd-GW}
\eea These generalize Eqs.\ (\ref{delta-eq-3rd})-(\ref{X-eq-3rd}) in
the presence of the gravitational waves. By combining Eqs.\
(\ref{delta-eq-3rd-GW}), (\ref{u-eq-3rd-GW}) we have \bea
   & & \ddot \delta + 2 H \dot \delta
       - 4 \pi G \mu \delta
   \nonumber \\
   & & \quad
       = - {1 \over a^2} {\partial \over \partial t}
       \left[ a \nabla \cdot \left( \delta {\bf u} \right) \right]
       + {1 \over a^2} \nabla \cdot \left( {\bf u}
       \cdot \nabla {\bf u} \right)
   \nonumber \\
   & & \quad
       + \dot C^{(t)\alpha\beta}
       \left( {2 \over a} u_{\alpha|\beta}
       + \dot C^{(t)}_{\alpha\beta} \right)
   \nonumber \\
   & & \quad
       + {1 \over a^2} {\partial \over \partial t}
       \left\{ a \left[ 2 \varphi {\bf u}
       - \nabla \left( \Delta^{-1} X \right) \right] \cdot \nabla \delta
       + 2 a \delta^{,\alpha} u^\beta C^{(t)}_{\alpha\beta}
       \right\}
   \nonumber \\
   & & \quad
       + {2 \over 3 a^2} \varphi
       {\bf u} \cdot \nabla \left( \nabla \cdot {\bf u} \right)
       - {4 \over a^2} \nabla \cdot \left[ \varphi
       \left( {\bf u} \cdot \nabla {\bf u}
       - {1 \over 3} {\bf u} \nabla \cdot {\bf u} \right) \right]
   \nonumber \\
   & & \quad
       + {\Delta \over a^2}
       \left[ {\bf u} \cdot \nabla \left( \Delta^{-1} X \right) \right]
       - {1 \over a^2} {\bf u} \cdot \nabla X
       - {2 \over 3a^2} X \nabla \cdot {\bf u}
   \nonumber \\
   & & \quad
       - {2 \over a^2} \left( \nabla \cdot {\bf u} \right)^{,\alpha} u^\beta
       C^{(t)}_{\alpha\beta}
       - {4 \over a} \dot C^{(t)\alpha\beta}
   \nonumber \\
   & & \quad
       \times
       \left[ u_\alpha \varphi_{,\beta}
       + 2 \varphi u_{\alpha|\beta}
       - {1 \over 2} \left( \Delta^{-1} X \right)_{,\alpha|\beta}
       + a \varphi \dot C^{(t)}_{\alpha\beta} \right]
   \nonumber \\
   & & \quad
       - 4 \left( {1 \over a} u^{\alpha|\beta}
       + \dot C^{(t)\alpha\beta} \right)
       \Bigg[ C^{(t)\gamma}_{\;\;\;\;\;\alpha}
       \left( {1 \over a} u_{\beta|\gamma}
       + \dot C^{(t)}_{\beta\gamma} \right)
   \nonumber \\
   & & \quad
       - {1 \over 3a} C^{(t)}_{\alpha\beta} \nabla \cdot {\bf u}
       + {1 \over 2 a} \left( 2 C^{(t)}_{\gamma\alpha|\beta}
       - C^{(t)}_{\alpha\beta|\gamma} \right) u^\gamma
       \Bigg],
   \label{density-eq-3rd-GW}
\eea which generalizes Eq.\ (\ref{density-eq-3rd}) to include the
gravitational waves.

Now, we present the equation for tensor-type perturbation to the
second order. {}From Eq.\ (103), (210) of \cite{NL} we can derive
equation for $\ddot C^{(t)}_{\alpha\beta}$ to the second order.
Since we are ignoring the vector-type perturbation, from Eqs.\
(211), (199) of \cite{NL} we have \bea
    & & \ddot C^{({t})}_{\alpha\beta}
        + 3 H \dot C^{({t})}_{\alpha\beta}
        - {\Delta \over a^2} C^{({t})}_{\alpha\beta}
        = N_{4\alpha\beta}
    \nonumber \\
    & & \qquad
        - {3 \over 2} \left( \nabla_\alpha \nabla_\beta
        - {1 \over 3} g^{(3)}_{\alpha\beta} \Delta \right)
       \Delta^{-2} \nabla^{\gamma} \nabla^\delta
       N_{4\gamma\delta}.
    \label{GW-eq}
\eea {}From Eq.\ (103) of \cite{NL} to the second order we have \bea
   & & N_{4 \alpha\beta}
       = {1 \over a^3} \Bigg\{ a^3 \Bigg[
       {2 \over a^2} \left( \varphi \chi_{,\alpha|\beta}
       + \varphi_{,(\alpha} \chi_{,\beta)} \right)
       + 2 \varphi \dot C^{(t)}_{\alpha\beta}
   \nonumber \\
   & & \quad
       + {2 \over a^2} \chi^{,\gamma}_{\;\;\;|\beta} C^{(t)}_{\alpha\gamma}
       + {1 \over a^2} \chi^{,\gamma}
       \left( 2 C^{(t)}_{\gamma(\alpha|\beta)}
       - C^{(t)}_{\alpha\beta|\gamma} \right)
   \nonumber \\
   & & \quad
       + 2 C^{(t)\gamma}_{\;\;\;\;\;\alpha} \dot C^{(t)}_{\beta\gamma}
       \Bigg] \Bigg\}^\cdot
   \nonumber \\
   & & \quad
       + {1 \over a^4} \chi^{,\gamma}_{\;\;\;|\alpha} \chi_{,\gamma|\beta}
       + {1 \over a^2} \left( \kappa \chi_{,\alpha|\beta}
       - 3 \varphi_{,\alpha} \varphi_{,\beta}
       - 4 \varphi \varphi_{,\alpha|\beta} \right)
   \nonumber \\
   & & \quad
       + \kappa \dot C^{(t)}_{\alpha\beta}
       + {1 \over a^2} \Bigg[
       2 \varphi^{,\gamma}_{\;\;\;|\alpha} C^{(t)}_{\beta\gamma}
       - 2 \Delta \varphi C^{(t)}_{\alpha\beta}
       - 4 \varphi \Delta C^{(t)}_{\alpha\beta}
   \nonumber \\
   & & \quad
       + \varphi^{,\gamma}
       \left( 2 C^{(t)}_{\gamma(\alpha|\beta)}
       - 3 C^{(t)}_{\alpha\beta|\gamma} \right)
       + 2 \chi^{,\gamma}_{\;\;\;|[\alpha} \dot C^{(t)}_{\beta]\gamma}
       - \chi^{,\gamma} \dot C^{(t)}_{\alpha\beta|\gamma}
   \nonumber \\
   & & \quad
       + 2 C^{(t)\gamma\delta} \left(
       2 C^{(t)}_{\gamma(\alpha|\beta)\delta}
       - C^{(t)}_{\alpha\beta|\gamma\delta}
       - C^{(t)}_{\gamma\delta|\alpha\beta} \right)
   \nonumber \\
   & & \quad
       - 2 C^{(t)\gamma}_{\;\;\;\;\;\alpha} \Delta C^{(t)}_{\beta\gamma}
       - C^{(t)\gamma}_{\;\;\;\;\;\delta|\alpha}
       C^{(t)\delta}_{\;\;\;\;\;\gamma|\beta}
       + 4 C^{(t)\gamma|\delta}_{\;\;\;\;\;\alpha}
       C^{(t)}_{\beta[\delta|\gamma]} \Bigg]
   \nonumber \\
   & & \quad
       - {1 \over 3} g^{(3)}_{\alpha\beta} \Bigg\{
       {1 \over a^3} \Big\{ a^3 \Big[
       {2 \over a^2} \left( \varphi \Delta \chi
       + \varphi^{,\gamma} \chi_{,\gamma} \right)
   \nonumber \\
   & & \quad
       + 2 C^{(t)\gamma\delta} \left( {1 \over a^2} \chi_{,\gamma|\delta}
       + \dot C^{(t)}_{\gamma\delta} \right) \Big] \Big\}^\cdot
       + {1 \over a^4} \chi^{,\gamma|\delta} \chi_{,\gamma|\delta}
   \nonumber \\
   & & \quad
       + {1 \over a^2} \Big[ \kappa \Delta \chi
       - 4 \varphi \Delta \varphi
       - 3 \varphi^{,\gamma} \varphi_{,\gamma}
       + 2 \varphi^{,\gamma|\delta} C^{(t)}_{\gamma\delta}
   \nonumber \\
   & & \quad
       - 4 C^{(t)\gamma\delta} \Delta C^{(t)}_{\gamma\delta}
       + C^{(t)\gamma\delta|\epsilon}
       \left( 2 C^{(t)}_{\gamma\epsilon|\delta}
       - 3 C^{(t)}_{\gamma\delta|\epsilon} \right) \Big]
       \Bigg\}.
   \label{N_4}
\eea In Eq.\ (\ref{N_4}) we have ignored $\alpha$ and $\dot \varphi$
terms which are already quadratic order in the comoving gauge. We
have $\chi = \chi_v$, $\varphi = \varphi_v$, $\kappa = \kappa_v$ and
$C^{(t)}_{\alpha\beta} = C^{(t)}_{\alpha\beta v}$ which are
gauge-invariant combinations. Apparently, we need $\chi_v$,
$\kappa_v$ and $\varphi_v$ to the linear order. To that order, we
have identified $\kappa_v = - {1 \over a} \nabla \cdot {\bf u}$ and
${\bf u} = {1 \over a} \nabla \chi_v$. {}For $\varphi_v$ we have
Eq.\ (\ref{varphi_v}). Using these identifications we can express
the scalar-type perturbation variables in Eq.\ (\ref{N_4}) in terms
of the Newtonian variables.

%
%
\section{Discussion}
                                              \label{sec:Discussions}

We have derived the third-order perturbation equations in the
zero-pressure cosmological medium in Einstein's gravity. We have
expressed the third-order terms using the Newtonian variables
identified in the lower order perturbations. Since the Newtonian
zero-pressure cosmological medium is exact to the second order in
perturbation, our third-order terms in relativistic context are pure
relativistic corrections. Our results show that the third-order
correction terms in relativistic energy and momentum conservation
equations are of $\varphi_v$-order higher than the second-order
terms, thus equivalently $\varphi_v$-order higher than the Newtonian
terms. The corrections terms are {\it independent} of the horizon
scale and depend only on $\varphi_v$ to the linear order which is
the spatial curvature perturbation in the comoving gauge
(hypersurface), or $\sim \delta \Phi$ which is the gravitational
potential perturbation strength. The variable $\varphi_v$ is known
to have conserved behavior and its amplitude in the large-scale
(near horizon, say) is constrained by the low-level anisotropies of
the CMB temperature, see Eq.\ (\ref{CMB-constraint}). Therefore, our
result reinforces our previous conclusion in \cite{second-order}
that one can use the large-scale Newtonian numerical simulation more
reliably even as the simulation scale approaches near (and goes
beyond) the horizon.

In this work we have {\it assumed} a single zero-pressure
irrotational fluid in the flat cosmological background. Dropping any
of these conditions could potentially lead to relativistic
corrections. Due to these assumptions we cannot apply our results
when the radiation components (including neutrino anisotropic
stress) become important in high redshift epoch, and in the case
when the baryon generated entropy leads to rotational perturbations
in the small-scale clusters.\footnote{We thank the anonymous referee
for making this point.} Extensions to include relativistic
second-order perturbational effects of the pressure, the rotation,
the non-flat background, and the multi-component situation will be
investigated in future occasions. In this work we derived the
relativistic correction terms appearing in the third order, and
showed that these correction terms do not involve the horizon scale
and are small in our observable patch.

Besides theoretical and practical significance we believe our exact
result to the second order and pure relativistic corrections to the
third order have historical value as well, because these have been
unsolved issues since Lifshitz's original work in the linear regime
in 1946 \cite{Lifshitz-1946}. Before our present third order and the
previous second order works there were different anticipations among
researchers in the field that even in the second order the
relativistic result might be different from the Newtonian ones: one
common anticipation was that the general relativistic effects might
become important as the scale approaches and goes beyond the
horizon. Our results resolved such an issue and showed that there
exist no correction terms to the second order in {\it all scales}.
We also showed that pure relativistic correction terms appearing in
third order do not depend on the horizon scale. It depends only on
the strength of the dimensionless gauge-invariant curvature
perturbation variable $\varphi_v$ or the dimensionless linear-order
gravitational potential $\delta \Phi$.

The post-Newtonian approximation \cite{Chandrasekhar-1965} provides
a complementary method to our perturbative approach in deriving the
relativistic correction terms in the Newtonian cosmology. The
post-Newtonian approximation takes $v/c$-expansion with $v/c \ll 1$,
thus for near virialized systems we have $GM/(Rc^2) \sim v^2/c^2 \ll
1$. Thus, although such an approximation takes into account of
nonlinearity it is valid only far inside the horizon; as we approach
the horizon $GM/(Rc^2)$ becomes unity. Consistency of the Newtonian
cosmology with the Newtonian limit of the post-Newtonian
approximation was reported in \cite{post-Newtonian}. We can show
that the Newtonian cosmological hydrodynamic equations naturally
appear in the zeroth-order post-Newtonian approximation \cite{HNP}.
Recently, we derived the fully nonlinear first-order post-Newtonian
correction terms, and showed that these correction terms have
typically $GM/(Rc^2) \sim v^2/c^2 \sim 10^{-5}$ order smaller than
the Newtonian terms in the non-linearly clustered regions
\cite{HNP}. Being a complimentary approach to the post-Newtonian
approximation (which provides fully non-linear equations) our
equations valid to the third-order perturbations may have diverse
applications in the cosmological situations where the systems have
not reached fully nonlinear stage. Compared with the post-Newtonian
approach, our perturbation approach is applicable in {\it all}
cosmological scale.

Even in the small (far less than the horizon) scale the pure
(third-order) relativistic correction terms could have important
roles if the strength of linear-order $\varphi_v$ is large enough.
Our relativistic results are valid in the perturbative sense. Thus,
if $\varphi_v$ approaches near unity, higher-order perturbative
terms could become important as well, and it is likely that our
perturbative approach breaks down. Still, it would be interesting to
investigate regimes where $\varphi_v$ is moderately important so
that we can study the roles of pure relativistic effects using our
third-order correction terms. {}For example, whether such pure
relativistic correction terms could lead to an observationally
distinguishable non-Gaussian signature \cite{non-Gaussian} is an
interesting issue which may deserve further attention. {}For such
investigations, Eqs.\ (\ref{delta-eq-3rd})-(\ref{density-eq-3rd})
are the complete set for pure scalar-type perturbation, and Eqs.\
(\ref{delta-eq-3rd-GW})-(\ref{N_4}) provide the complete set
including the gravitational waves. As we consider a flat background
the ordinary Fourier analysis can be used to study the
mode-couplings as in the Newtonian case in \cite{quasilinear}.
Comparing the roles of pure relativistic third-order corrections
with the inherent third-order perturbation effects in the Newtonian
approximation in Eqs.\ (\ref{delta-eq-3rd})-(\ref{density-eq-3rd})
will be an interesting step we can take. Applications to such
cosmological situations are left for future studies.

%
%
\subsection*{Acknowledgments}

HN and JH were supported by grants No. R04-2003-10004-0 and
No. R02-2003-000-10051-0, respectively, from the
Basic Research Program of the Korea Science and Engineering Foundation.



\begin{thebibliography}{99}
\bibitem{NL}
         H. Noh and J. Hwang, Phys. Rev. D {\bf 69}, 104011 (2004).
\bibitem{second-order}
         H. Noh and J. Hwang, Class. Quant. Grav. {\bf 22}, 3181 (2005);
         J. Hwang and H. Noh, Phys. Rev. D {\bf 72}, 044011 (2005).
\bibitem{Friedmann-1922}
         A.A. Friedmann, Zeitschrift f\"ur Physik {\bf 10}, 377 (1922),
                         and
              {\it ibid.} {\bf 21}, 326 (1924);
              both papers are translated in
                         {\it Cosmological-constants: papers in modern
                         cosmology}, edited by J. Bernstein and G. Feinberg
                         (Columbia Univ. Press, New York, 1986), p49 and p59;
         H.P. Robertson, Proceedings of the National Academy of Science
                         {\bf 15}, 822 (1929).
\bibitem{Milne-1934}
         E.A. Milne, Quart. J. Math. {\bf 5}, 64 (1934);
         W.H. McCrea and E.A. Milne, {\it ibid.} {\bf 5}, 73 (1934).
\bibitem{Lifshitz-1946}
         E.M. Lifshitz, J. Phys. (USSR) {\bf 10}, 116 (1946).
\bibitem{Bonnor-1957}
         W.B. Bonnor, Mon. Not. R. Astron. Soc. {\bf 117}, 104 (1957).
\bibitem{Peebles-1980}
         P.J.E. Peebles, {\it The large-scale structure of the universe},
                         (Princeton Univ. Press, Princeton, 1980).
\bibitem{HN-Newtonian-1999}
         J. Hwang and H. Noh, Gen. Rel. Grav. {\bf 31}, 1131 (1999).
\bibitem{observations}
         D.N. Spergel, {\it et al.}, Astrophys. J. Suppl. {\bf 148}, 175 (2003);
         M. Tegmark, {\it et al.}, Phys. Rev. D {\bf 69}, 103501 (2004).
\bibitem{Bardeen-1988}
         J.M. Bardeen, {\it Particle Physics and Cosmology}, edited by
                       L. Fang and A. Zee (Gordon and Breach, London, 1988), p1.
\bibitem{H-1991}
         J. Hwang, Astrophys. J. {\bf 375}, 443 (1991).
\bibitem{covariant}
         J. Ehlers, Proceedings of the mathematical-natural science of
                    the Mainz academy of science and literature,
                    Nr. {\bf 11}, 792 (1961),
                    translated in Gen. Rel. Grav. {\bf 25}, 1225 (1993);
         G.F.R. Ellis, in {\it General relativity and cosmology, Proceedings of
                       the international summer school of physics Enrico
                       Fermi course 47}, edited by R. K. Sachs (Academic
                       Press, New York, 1971), p104;
                       in {\it Cargese Lectures in Physics}, edited by
                       E. Schatzmann (Gorden and Breach, New York, 1973), p1.
\bibitem{ADM}
         R. Arnowitt, S. Deser, and C.W. Misner, in {\it Gravitation: an
                      introduction to current research}, edited by  L. Witten
                      (Wiley, New York, 1962) p. 227.
\bibitem{Bardeen-1980}
         J.M. Bardeen, Phys. Rev. D {\bf 22}, 1882 (1980).
\bibitem{H-MDE-1994}
         J. Hwang Astrophys. J. {\bf 427}, 533 (1994).
\bibitem{SW-1967}
         R.K. Sachs and A.M. Wolfe, Astrophys. J. {\bf 147}, 73 (1967).
\bibitem{HN-SW1}
         J. Hwang and H. Noh, Phys. Rev. D {\bf 59}, 067302 (1999).
\bibitem{CMB}
         G.F. Smoot, {\it et al.} Astrophys. J. {\bf 396}, L1 (1992).
\bibitem{conserved}
         J. Hwang, Phys. Rev. D {\bf 53}, 762 (1996);
                   J. Korean Phys. Soc. {\bf 35}, S633 (1999);
         J. Hwang and H. Noh, Phys. Rev. D {\bf 61}, 043511 (2000);
                  {\it ibid.} {\bf 66}, 084009 (2002);
                  {\it ibid.} {\bf 71}, 063536 (2005).
\bibitem{Salopek-Bond-1990}
         D.S. Salopek and J.R. Bond, Phys. Rev. D {\bf 42}, 3936 (1990).
\bibitem{Chandrasekhar-1965}
         S. Chandrasekhar, Astrophys. J. {\bf 142}, 1488 (1965).
\bibitem{post-Newtonian}
         E. Bertschinger and A.J.S. Hamilton,
                         Astrophys. J. 435 (1994) 1;
         L. Kofman and D. Pogosyan, {\it ibid.} 442 (1995) 30;
         G.F.R. Ellis and P.K.S. Dunsby, Astrophys. J. {\bf 479}, 97
                      (1997).
\bibitem{HNP}
         J. Hwang, H. Noh, and D. Puetzfeld, Phys. Rev. D submitted (2005),
                   astro-ph/0507085.
\bibitem{non-Gaussian}
         F. Bernardeau, S. Colombi, E. Gaztanaga, and R. Scoccimarro,
                        Phys. Rep. {\bf 367}, 1 (2002);
         N. Bartolo, E. Komatsu, S. Matarrese, and A. Riotto,
                     Phys. Rep. {\bf 402}, 103 (2004).
\bibitem{quasilinear}
         E.T. Vishniac, Mon. Not. R. Astron. Soc. {\bf 203}, 345 (1983);
         M.H. Goroff, B. Grinstein, S.-J. Rey, and M.B. Wise,
                      Astrophys. J., {\bf 311}, 6 (1986).
\end{thebibliography}
\end{document}